\documentclass[prl,aps,twocolumn,floats,nofootinbib]{revtex4-1}
\usepackage{epsfig,bm,dcolumn}
\usepackage{graphicx}
\usepackage{ulem}
\usepackage{epsfig}
\usepackage{latexsym}
\usepackage{bm}
\usepackage{amsmath,amssymb,graphicx,psfrag}
\usepackage[colorlinks,linkcolor={blue},citecolor={blue},urlcolor={blue}]{hyperref}
\usepackage{caption}
\usepackage{xcolor}

\begin{document}
\renewcommand{\ni}{{\noindent}}
\newcommand{\dprime}{{\prime\prime}}
\newcommand{\be}{\begin{equation}}
\newcommand{\ee}{\end{equation}}
\newcommand{\bea}{\begin{eqnarray}} 
\newcommand{\eea}{\end{eqnarray}}
\newcommand{\la}{\langle}
\newcommand{\ra}{\rangle} 
\newcommand{\dg}{\dagger}
\newcommand\lbs{\left[}
\newcommand\rbs{\right]}
\newcommand\lbr{\left(}
\newcommand\rbr{\right)}
\newcommand\f{\frac}
\newcommand\e{\epsilon}
\newcommand\ua{\uparrow}
\newcommand\da{\downarrow}
\newcommand\mbf{\mathbf}

\newcommand{\Eqn}[1] {Eqn.~(\ref{#1})}
\newcommand{\Fig}[1]{Fig.~\ref{#1}}

\title{Universal properties of single particle excitations across the many-body localization transition}
\author{Atanu Jana$^{1,3}$}\author{V. Ravi Chandra$^{1,3}$}\author{Arti Garg$^{2,3}$} 
\affiliation{$^{1}$ School of Physical Sciences, National Institute of Science Education and Research, Bhubaneswar, Jatni, Odisha 752050, India}
\affiliation{$^{2}$ Theory Division, Saha Institute of Nuclear Physics, 1/AF Bidhannagar, Kolkata 700 064, India}
\affiliation{$^{3}$ Homi Bhabha National Institute, Training School Complex, Anushaktinagar, Mumbai 400094, India}
\vspace{0.2cm}
\begin{abstract}
  \vspace{0.3cm}
  Understanding the nature of the transition from the delocalized to the many-body localized (MBL) phase is an important unresolved issue. To probe the nature of the MBL transition, we investigate the universal properties of single-particle excitations produced in highly excited many-body eigenstates of a disordered interacting quantum many-body system. In a class of one-dimensional spinless fermionic models with random disorder,  we study the finite size scaling of the ratio of typical to average values of the single-particle local density of states and the scattering rates across the MBL transition. Our results indicate that the MBL transition in this class of one-dimensional models of spinless fermions is continuous in nature.  For various ranges of interactions in the system, the critical exponent $\nu$ with which the correlation length $\xi$ diverges at the transition point $W_c$, $\xi \sim |W-W_c|^{-\nu}$, satisfies the Chayes-Chayes-Fisher-Spencer(CCFS) bound $\nu \ge 2/d$ where $d$ is the physical dimension of the system. We also discuss why the critical exponent obtained from finite-size scaling of the conventional diagnostic of many-body localization, the level-spacing ratio, strongly violates the CCFS bound while the single-particle density of states and scattering rates are consistent with the CCFS criterion.

\vspace{0.cm}
\end{abstract} 
\maketitle

\section {I. Introduction}
The role of disorder in quantum many-body systems has been a major focus of research in condensed matter physics for several decades. Anderson localization is an astonishing example of a disorder-driven phenomenon in which a non-interacting quantum system can become diffusion-less in the presence of strong enough disorder~\cite{Anderson}. Almost two decades ago, Anderson localization  was generalised for the case of interacting quantum systems ~\cite{Basko,Mirlin_0} which is known as  many-body localization (MBL)~\cite{MBL_rev}. In the MBL phase, a subsystem of an isolated quantum system does not thermalize with the rest of the system serving as its bath ~\cite{MBL_rev,Huse_2007,Luitz} and the system has strong memory of initial states~\cite{Alet,expt1,expt2,expt3,Gornyi,Mirlin_imb,Titas,yp_nee,Popperl,yp}. 
  Even highly excited states of an isolated MBL system obey area law of entanglement entropy~\cite{MBL_rev,Luitz} and the system has a slow growth of the subsystem entanglement in a quench protocol~\cite{Moore,Serbyn_EE,yp}. Although the MBL phase has been rigorously proved to exist in strongly disordered 1-dimensional spin chains with short range interactions~\cite{Imbrie}, broad agreement about the nature of the transition from the delocalized phase to the MBL phase has been elusive. We provide strong evidence in favor of a continuous transition from the delocalized phase to the MBL phase in this work.

  The MBL transition is an atypical transition which does not necessarily follow the standard paradigm used to classify phase transitions. It is not easy to identify the local order parameters that can characterize the delocalization to MBL transition. This makes it crucial to search for criteria that can provide hints towards the nature of the MBL transition. One such criterion is given by  Chayes-Chayes-Fisher-Spencer(CCFS) bound on the critical exponent $\nu$ with which the correlation length $\xi$ diverges at the transition point~\cite{CCFS}. According to the CCFS criterion, for all systems with quenched random disorder that undergo a continuous transition including Anderson localization transition, $\nu \ge 2/d$, where $d$ is the physical dimension for the system, irrespective of whether there is an analogous transition in the clean system~\cite{CCFS,footnote}. In fact, the finite size scaling of the Anderson localization transition for the non-interacting model ($d \ge 3$) has been shown to satisfy the CCFS bound for the critical exponent~\cite{Boris,Kramer,Slevin,Biroli,Huse_AM} and one would expect it to hold true even for the MBL transition. 

In the context of MBL, some phenomenological real-space renormalization group studies predicted a critical point at the MBL transition with the critical exponent $\nu \sim 3$~\cite{RG_Ehud,RG_Potter,RG_zhang,RG_Dumit,RG_Yao} that satisfies the CCFS bound. 
One major source of concern has been that the finite-size scaling analysis for the conventional characterizations of the MBL phase, such as level spacing ratio and entanglement entropy give the critical exponent $\nu \le 1$ violating the CCFS bound~\cite{Bardarson,Luitz,khemani,Piotr}. There are only a few exceptions, such as the Schmidt gap, which has been shown to be consistent with the CCFS criterion~\cite{RIM,LRME,Bayat2}. 
The violation of the CCFS bound, as well as the disparity  between phenomenology and numerical calculations prompted an avalanche based~\cite{avalanche,morningstar2} renormalization group approach ~\cite{Vasseur1,Vasseur2,morningstar1} that predicted a Kosterlitz-Thouless (KT) like transition and has been explored in some recent numerical studies~\cite{Prosen,Mace,Meisner}.
In short, there is no agreement on the nature of the delocalization to MBL transition, so it is essential to identify appropriate physical observables that can characterize the MBL transition.

With this motivation, in this work we investigate single-particle excitations obtained via single-particle Green's functions in real space calculated in highly excited many-body eigen-states across the MBL transition. Green's functions in real space have been widely utilised to analyse Anderson localization in non-interacting models~\cite{AL_book} but single particle Green's functions have only recently received attention in the analysis of the MBL phase~\cite{Atanu}. 
We analyse the finite size scaling of the local density of states (LDOS) and the scattering rates. 
We demonstrate that the ratio of the typical to average value of the local density of states as well as the scattering rates both indeed adhere to the single parameter scaling $X[L,W] \sim \bar{X}((W-W_C)L^{1/\nu})$ with the critical exponent satisfying the CCFS inequality for a finite value of $W_c$. Notably,  we observe a good quality scaling collapse with $\nu \ge 2/d$ for the ratio of the typical to average value of the LDOS as well as the scattering rates not only for the system with nearest neighbour interactions but also for a whole class of one dimensional models with power-law interactions of different ranges and nearest neighbour hopping. Though generally a power-law diverging correlation length at the transition point is associated with a continuous transition, for a dynamical transition like the MBL transition this may not always be true~\cite{Roeck}. But the physical quantities that we explore in this work, namely, single-particle local density of states and their scattering rates, seem to decay continuously at the MBL transition and also satisfy the CCFS criterion. Therefore, we will associate the term continuous transition with the CCFS criterion in this work which is also consistent with the nomenclature in some of the earlier renormalization group studies on MBL~\cite{RG_Ehud,RG_Potter,RG_zhang,RG_Dumit,RG_Yao}, numerical works on MBL transition~\cite{Bardarson,Luitz,khemani,Piotr} as well as with the original CCFS paper. Finite size scaling of eigenlevel spacing ratio, on the other hand, does not satisfy the CCFS bound of the critical exponent which is consistent with earlier studies~\cite{Bardarson,Luitz,khemani,Piotr}.

The rest of the paper is organised as follows. In section (II) we describe the class of models investigated in this work and the method used to analyse the MBL transition. In section (III) we present the results of our numerical analysis and the details of the finite-size scaling. Finally we conclude with discussion on open questions and subtle issues in section (IV). 

\section {II. Model and Method}
We study a class of one-dimensional models of spinless fermions in the presence of random disorder and power-law interactions.  The Hamiltonian of the models studied is
\bea
H=-t\sum_{i}[c^\dagger_ic_{i+1}+h.c.] + \sum_i \epsilon_i n_i + \sum_{ij}V_{ij}n_in_{j}
\label{Ham}
\eea
with periodic boundary conditions. Here, the onsite potential $\epsilon_i \in [-W/t,W/t]$ (uniformly distributed) with $W$ as the disorder strength. 
We study power-law interactions with $V_{ij}=\frac{V}{|r_i-r_j|^\alpha}$, where $\alpha$ fixes the range of interactions. We have considered
$\alpha=1, 2$ and $3$ in this study. We also consider the limit of the very short range interactions by studying  
the case of nearest neighbour interactions with $V_{i,i+1}=V$ and $V_{ij}=0$ for $|j-i| > 1$. 
In the entire analysis the strength of interactions has been fixed to be $V=t(=1)$ and the system is half-filled. We  study the model using exact diagonalization, for several system sizes from $L=12$ to $L=18$. For each value of $\alpha$ we use $[15000-50]$ realisations of disorder for $L=[12-18]$  to calculate the averages of the LDOS and scattering rates. For the system with nearest neighbour interactions, we use $[15000-200]$ realizations of disorder for $L=[12-18]$ respectively.

We study the Green's function in the $nth$ eigenstate $ G_{n}(i,j, \tau) = -\iota \Theta(\tau) \langle \Psi_n  \vert \{ c_i(\tau), c_j^\dg(0)\} \vert \Psi_n \rangle$, where $i,j$ are lattice site indices and $\tau $ is real time. Fourier transform of the Green's function to frequency space results in the Lehmann representation of $G_n(i,j,\omega)$ as shown below:
\be
G_n(i,i,\omega^+) = \sum_m \frac{|\la\Psi_m|c^\dagger_i|\Psi_n\ra|^2}{\omega+i\eta-E_m+E_n} +\frac{|\la\Psi_m|c_i|\Psi_n\ra|^2}{\omega+i\eta+E_m-E_n}
\label{Gn}
\ee
The associated self energy is ${\boldsymbol{\Sigma}_n} (\omega) \equiv \mbf{{{{G^{-1}_{0}(\omega) - G^{-1}_{n}(\omega)}}}}$ where $\mbf{G_0}(\omega)$ and $\mbf{G_n}(\omega)$ are Fourier transforms of the non-interacting and interacting Green's function matrices respectively. The LDOS $\rho_{n}(i, \omega)$ and scattering rate are obtained from the imaginary part of the Green's function and the self energy respectively as $\rho_{n}(i, \omega) = \left( - \frac{1}{\pi} \right ) Im \left [ G_{n}(i, i, \omega+\iota \eta) \right)$ and ${\Gamma}_n(i, \omega) = - Im \left[\Sigma_n (i,i,\omega+\iota\eta)\right]$. The broadening $\eta$  should be of the order of but larger than the typical spacing between the adjacent eigenvalues for all the parameters considered in the study~\cite{Parisi,Zhu,Wu,Bari,Berkelbach}. In the thermodynamic limit, in the localized phase the typical value of the LDOS scales proportionally to $\eta$ while in the delocalized phase the typical LDOS is independent of $\eta$. For a finite size system, this independence of typical LDOS in the delocalized phase is seen for a range of $\eta$ between the average value of the level spacing of a system of size $L$ and the average level spacing of the system of size equal to the correlation length and in the thermodynamic limit, the two length scales merge approaching zero~\cite{eta_Mirlin,eta_Altshuler,Altshuler2,eta_Brezini}. We followed this approach and explored the $\eta$ dependence of the typical DOS, details of which are provided in Appendix A. By checking representative values of disorder and system sizes, we found that for $0.0075 \le \eta \le 0.03$, typical value of LDOS is independent of the broadening $\eta$ in the delocalized phase and we presented the results for $\eta=0.01$.

 \begin{figure}
   \begin{center}
     \hspace{-1cm}
        \includegraphics[width=3.15in,angle=0]{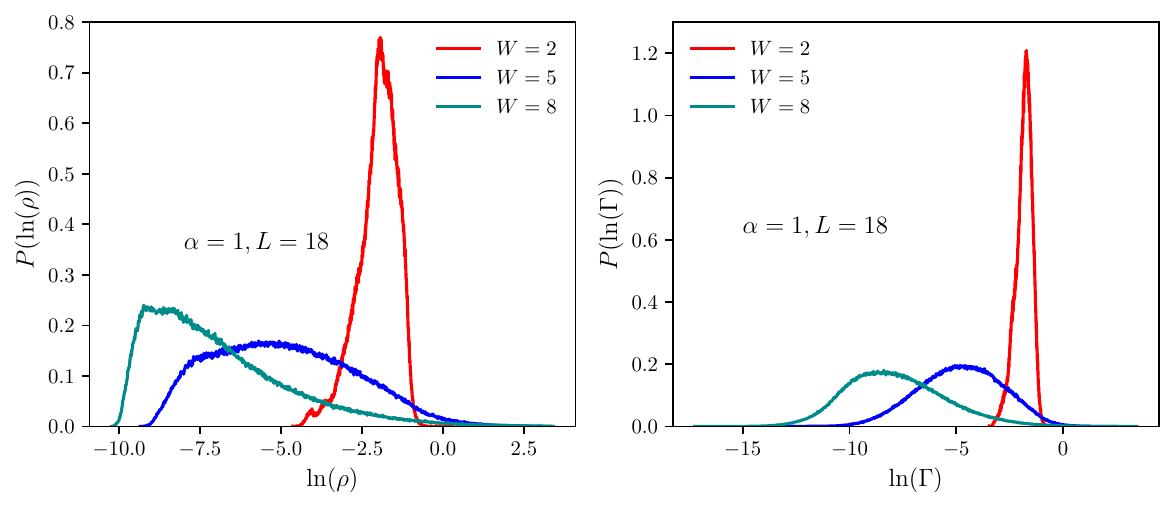}
          \caption{ Probability distribution function of the logarithm of the local density of states $\rho(\omega\sim \mu_{eff})$ for a few values of disorder $W$. For weak disorder, $P(\ln(\rho))$ is close to a normal distribution. As the disorder strength increases, the peak of the distribution shifts towards smaller values and the width of the distribution increases. The data shown is for power-law interacting system with $\alpha=1$. The right panel shows the probability distribution function $P(\ln(\Gamma(\mu_{eff})))$ for the scattering rates.}
          \label{prob}
          \end{center}
  \end{figure}

 The transition from the delocalized to the MBL phase is seen in the disordered averaged Green's function calculated for the mid spectrum eigen-states with rescaled energy $\epsilon_n =\frac{E_n-E_{min}}{E_{max}-E_{min}}\sim 0.5$. This is because many-body states in the middle of the spectrum require largest strength of disorder to get localized~\cite{MBL_rev}, and if one restricts the analysis only to the ground state, as was done in several earlier works for higher dimensional disordered interacting systems~\cite{int_AL1,int_AL2}, it would not be possible to capture the physics of the MBL transition. Further,  we analyse the ratio of typical to average value of the LDOS and scattering rates. Here, the typical value is obtained by calculating the geometric average over the lattice sites, energy bin and various independent disorder configurations. 

 \section{III. Results}
 
 In this section we present results for LDOS and scattering rates for various range of interactions in the system. For systems with power-law interactions typical value of local density of states has a peak around $\omega=\mu_{eff}$ where $\mu_{eff}= \sum_{|j-i|=1}^{L/2} \frac{1}{|j-i|^{\alpha}}$ (for $V=1$) is the effective chemical potential of the system under the assumption that the disorder averaged system will respect particle hole symmetry (Details in Appendix B). Thus, for power-law interacting case we have shown scaling of ratio of typical to average value of LDOS and scattering rate for $\omega \sim \mu_{eff}$. For the system with nearest neighbour interactions we have shown results for $\omega=0$. The LDOS $\rho_{typ}(\omega)$ and the scattering rate $\Gamma_{typ}(\omega)$ are very flat around $\omega=0$ over a width of around $2W$ and $\mu_{eff}$ for nearest neighbour interaction is $V/2$. Thus, effectively the behaviour of LDOS and scattering rate at $\omega=\mu_{eff}$ and $\omega=0$ is almost the same for the system with nearest neighbour interactions.
 \\\\
 \noindent{\bf{Probability Distribution Functions}}:
 We first look at the probability distribution functions for the LDOS and scattering rates. Fig.~\ref{prob} shows the probability distribution function of $\ln(\rho(\mu_{eff}))$ for various values of disorder $W$ for the power-law interacting system with $\alpha=1$. For weak disorder $P(\ln(\rho))$ is close to a normal distribution, that is, LDOS obeys the log-normal distribution but for larger values of disorder, the distribution deviates from log-normal distribution significantly. This shows that both the physical quantities under consideration in this disordered interacting system are asymmetrically distributed with long tails and the typical value is a more appropriate distinguishing characteristic of such distributions. This is in analogy to the non-interacting Anderson model, where the typical value of the local density of states acts as the order parameter across the localization transition rather than its average value~\cite{AL_book,Janssen}.
As the disorder strength increases, the peak of  $P(\ln(\rho))$ shifts to more negative values and the tail becomes broader. This is reflected in smaller values of the typical LDOS compared to the average value of the distribution as $W$ increases. Right panel of Fig.~\ref{prob} shows the probability distribution of $\ln(\Gamma(\mu_{eff}))$ which is closer to log-normal distribution even for larger values of the disorder strength.
\\
\\
\begin{figure*}[t]
\centering
\includegraphics[scale=0.45]{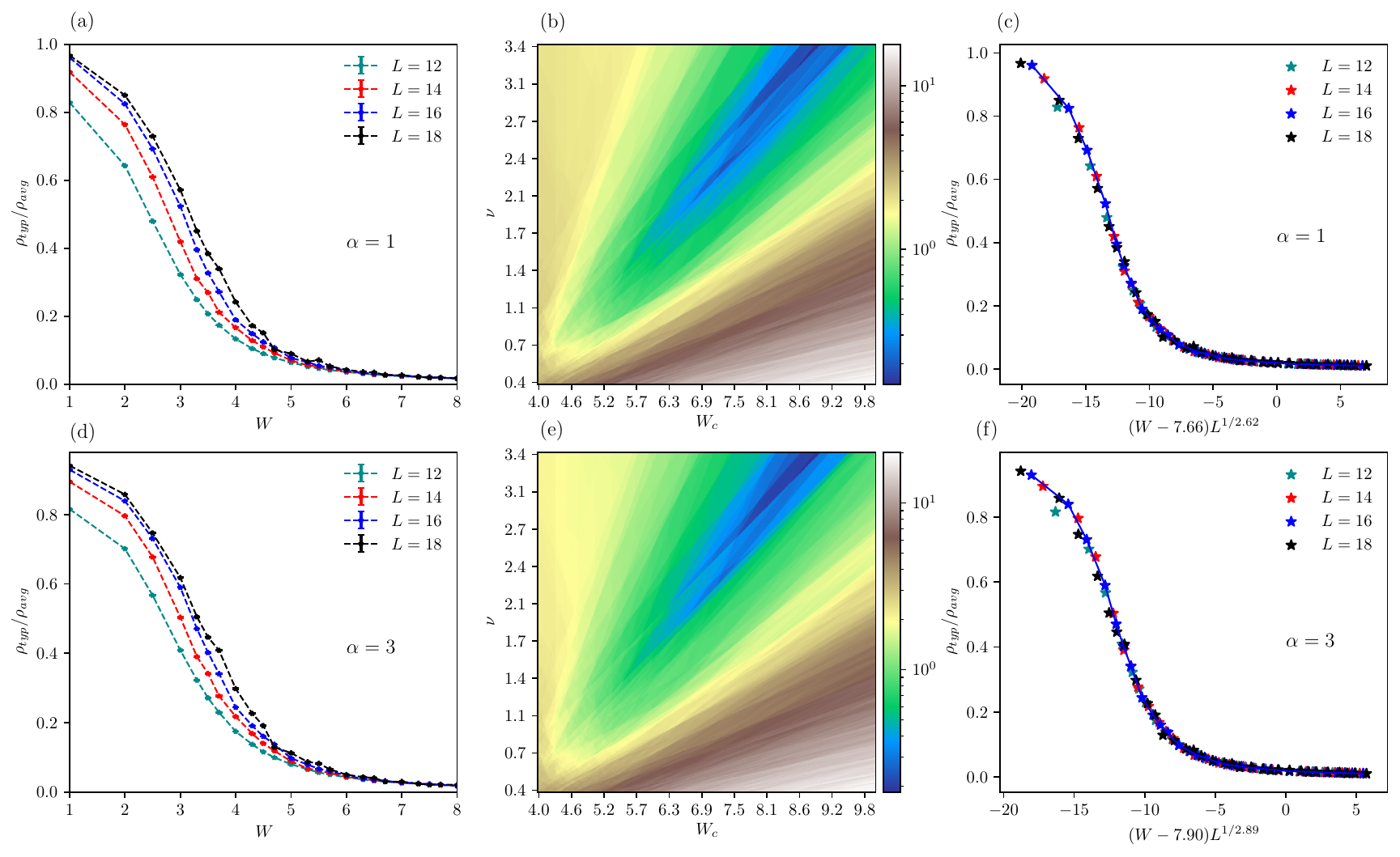}
\caption{Panel (a): The ratio of the typical to average local DOS $\rho_{typ}(\omega=\mu_{eff})/\rho_{avg}(\omega=\mu_{eff})$ as a function of the disorder strength $W$ for $\alpha=1$. The ratio is of order one for $W \ll W_c$ and for $W > W_c$, it is vanishingly small.  Panel (b): The cost function $C_X$ (in Eq.~\ref{cost}) for $X=\rho_{typ}(\omega=\mu_{eff})/\rho_{avg}(\omega=\mu_{eff})$ as a function of the critical disorder strength $W_c$ and the correlation length exponent $\nu$. Panel (c):  
The ratio of the typical to average value of local DOS $\rho_{typ}(\omega=\mu_{eff})/\rho_{avg}(\omega = \mu_{eff})$ plotted 
as a function of scaled disorder strength  $(W-W_c)L^{1/\nu}$ for $W_c= 7.66t$ and $\nu = 2.62$ corresponding to the region of the cost function shown in the middle panel with the minimum value of $C_x$. The bottom three panels
depict the same quantities for $\alpha=3$. The calculations are done in the middle of the energy band for a rescaled energy bin $\epsilon \in [0.495,0.505]$.} 
\label{Fig1}
\end{figure*}
\noindent{\bf{Typical LDOS and scattering rates for power-law interactions}}:
In the weak disorder limit, for any range of interactions, single-particle excitations are extended. 
Panel (a) of Fig.~\ref{Fig1} shows the ratio of the typical to average value of the LDOS $\rho(\omega)$ for the system with power-law interactions with $\alpha=1$. 
For weak disorder, the typical value of the LDOS is of the order of the average LDOS while for large values of $W$ in the MBL phase the typical value of the LDOS becomes vanishingly small for all values of $\omega$ (as shown in Appendix B) though the corresponding average value is still finite. The ratio of typical to average value of LDOS increases with the system size for weak disorder while for very large disorder it becomes essentially independent of the chain size.  Interestingly, at the disorder value $W^\star \sim 7.1t$ where the ratio shows very weak dependence on the system size, it also becomes constant with respect to disorder $W$ within numerical precision.  Thus, the single-particle excitations capture the basic features of delocalization to MBL transition even for systems with various ranges of interactions. Below we perform the finite-size scaling of the ratio of LDOS and scattering rates in order to investigate the nature of the MBL transition. 

\vspace{1cm}
\noindent{\bf{Finite-size Scaling Analysis}}: We assume that the characteristic length scale diverges with a power law at the MBL transition point $\xi \sim |W-W_C|^{-\nu}$. As a result a normalized observable $X$ obeys the scaling $X[\delta,L] \sim \bar{X}(\delta L^{1/\nu})$ with $\delta=W-W_c$. To have a quantitative estimate of the scaling collapse, we calculate the cost-function for the quantity $\{X_i\}$~\cite{Prosen,Somen}. 
\be
C_X=\frac{\sum_{j=1}^{N_{total}-1}|X_{j+1}-X_j|}{max\{X_j\}-min\{X_j\}} -1
\label{cost}
\ee
Here $N_{total}$ is the total number of values of $\{X_i\}$ for various values of disorder $W$ and system sizes $L$. We arrange all $N_{total}$ values of $\{X_i\}$ according to increasing values of $(W-W_C)L^{1/\nu}$. Ideally $C_X$ should be zero for a perfect data collapse but for the finite size data that we have, we look for a minimum of the cost function in $(W_c,\nu)$ plane.
We study the ratios of typical to average LDOS and scattering rates introduced earlier using a single parameter scaling form $(X[\delta,L] \sim \bar{X}(\delta L^{1/\nu}))$, 
which has also been used to study scaling properties of other quantities relevant in context of MBL~\cite{Luitz,Bardarson,khemani,Piotr}. As we will show shortly,  
this scaling ansatz results in very good scaling collapse for these quantities.

The computation of the cost function has numerical uncertainties which are inherited from the errors in our raw data. These errors in the cost function evaluation can be determined using standard error propagation methods. They can be used to obtain an estimate and uncertainty of both the critical parameters. Furthermore, we can also estimate the confidence intervals of the parameters by developing a bootstrap-like resampling method of our data by using subsets of disorder configurations for each length scale. In Appendix D we provide details of such analyses using both these approaches along with the details of error estimation and minimization of the cost function. 


The cost function for the ratio of typical to average values of the LDOS is shown in panel (b) of Fig.~\ref{Fig1} for $\alpha=1$. $C_X$ decreases as the value of the parameter $W_c$ is increased from $5t$, having a minimum around $7.67t$ and $\nu \sim 2.62$.  
With further increase in $W_c$ and $\nu$, $C_X$ shows a slow increase. 
The finite size scaling collapse shown in panel (c) of Fig.~\ref{Fig1} has been made for $W_c=7.67t$ and $\nu = 2.62$ though any point in $(W_c,\nu)$ plane corresponding to the minimum region of the cost function within error-bar would give a good quality scaling collapse (details in Appendix D). 
The bottom panel in Fig.~\ref{Fig1} shows similar plots for $\alpha=3$. As one can see that the finite size scaling and the minimization of the cost function provides a critical point $W_c \sim 7.90t$ which is slightly larger than the  transition point obtained for $\alpha=1$. This is consistent with earlier works on disordered spin chains with long range $ZZ$ couplings~\cite{Logan,Bayat}. The critical exponent $\nu \in [2.87,2.90]$ with $95\%$ confidence interval (Details in Appendix D). In fact, the critical exponent $\nu$ continues to satisfy the CCFS criterion even for $\alpha=0.5$ (results not shown here). We also studied the finite size scaling of the ratio of typical to average value of scattering rates and obtain the critical exponent $\nu>2$ satisfying the CCFS criterion as that from the LDOS, details of which are provided in Appendix C.

\begin{figure}[h!]
  \begin{center}
    \hspace{-0.6cm}
\includegraphics[scale=0.35]{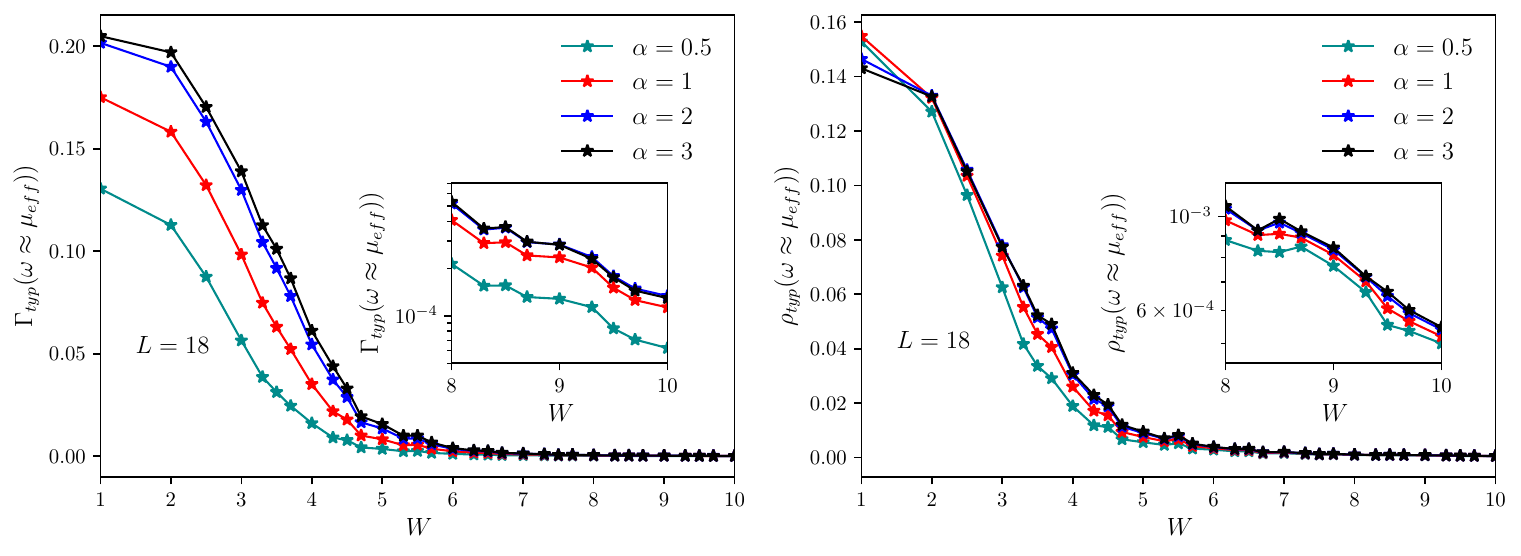}
\caption{Panel [a]: Typical values of scattering rates $\Gamma_{typ}(\mu_{eff})$ vs $W$ for a range of $\alpha$ values. For a fixed disorder strength $W$, as the range of interactions increases, $\Gamma_{typ}(\mu_{eff})$ decreases indicating more localized single-particle excitations for smaller $\alpha$ values. Panel [b]: Typical values of LDOS $\rho_{typ}(\mu_{eff})$ vs $W$ for various $\alpha$ values. All quantities are shown for $L=18$ and are computed for states in the middle of the eigenspectrum for a rescaled energy $\epsilon \in [0.495,0.505]$.}

\label{alpha}
\end{center}
\end{figure}

For a fixed disorder strength as the range of interaction increases, $\rho_{typ}(\mu_{eff})$ and $\Gamma_{typ}(\mu_{eff})$ decrease indicating enhanced effect of disorder and stronger tendency towards localization for smaller values of $\alpha$ as shown in Fig~\ref{alpha}. This fascinating effect of the range of interactions is seen in LDOS and scattering rates at all frequencies as shown in Appendix B. This is consistent with earlier observations of enhanced tendency towards localization in terms of increased return probability and reduced density imbalance in the long time limit of long range interacting fermionic MBL systems~\cite{garg_lr}. This is also qualitatively consistent with the studies on long range ZZ coupling in disordered spin chains~\cite{Burin,Sarang,Gutman,Mirlin,Logan,Bayat}.
We would like to emphasize, that despite the fact that we do observe signatures of a more stable MBL phase in the presence of longer range interactions in terms of more localized single-particle excitations, one should proceed with caution when making inferences regarding the slight drift of the transition point $W_c$ towards smaller values with increase in the range of interactions. Since in exact diagonalization there is a limitation of the maximum system size that can be studied, finite-size scaling over these systems sizes can not provide transition point with better precision especially for systems with long-range interactions. Nevertheless, our results are in qualitative consistency with earlier studies on long range MBL systems~\cite{Burin,Sarang,Gutman,Mirlin,garg_lr,Bayat}. At this point, we would like to mention that there are issues related with stability of the MBL phase in the presence of long-range interactions arising due to the presence of rare thermal bubbles in systems with random disorder~\cite{bubble}. But these effects are significant for $\alpha \ge 2d$ as discussed in detail earlier~\cite{Mirlin}.
\begin{figure*}[t]
  \begin{center}
  \hspace{-1cm}
\includegraphics[scale=0.45]{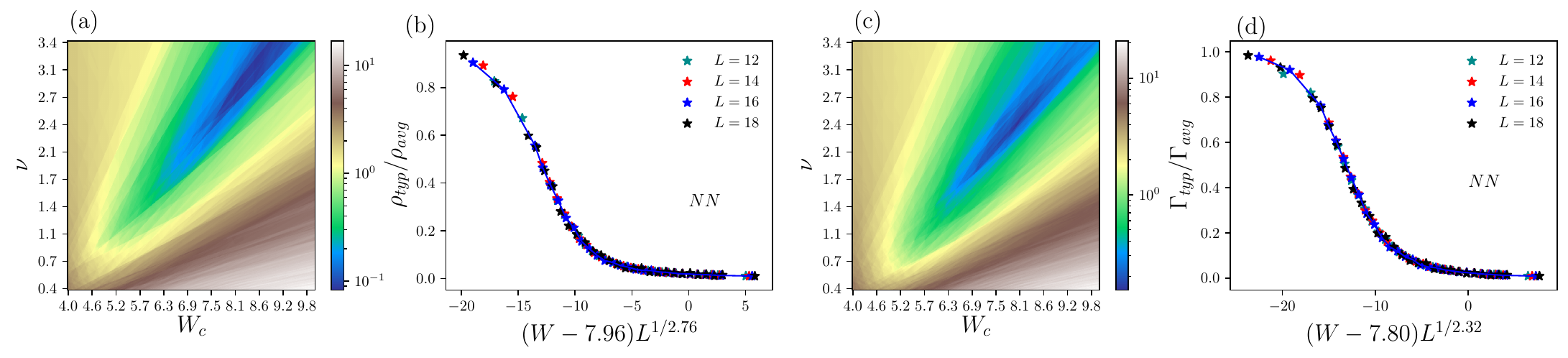}
\caption{Panel [a]: The cost function in $W_c-\nu$ plane for the ratio of the typical to average LDOS for nearest neighbour interacting system. Panel [b]: The ratio of the typical to average LDOS $\rho_{typ}(\omega=0)/\rho_{avg}(\omega=0)$ plotted as a function of the scaled disorder strength $(W-W_c)L^{1/\nu}$. The critical disorder $W_c \sim 7.96t$ and the exponent $\nu \sim 2.76$ are obtained by minimising the cost function $C_X$ (in Eq.~\ref{cost}). Panel [c]:  The cost function in $W_c-\nu$ plane for the ratio of the typical to average scattering rates for nearest neighbour interacting system. Panel [d] : The ratio of the typical to average value of the scattering rate $\Gamma_{typ}(\omega=0)/\Gamma_{avg}(\omega=0)$ as a function of the scaled disorder $(W-W_c)L^{1/\nu}$. All quantities are computed for states in the middle of the eigenspectrum for a rescaled energy $\epsilon \in [0.495,0.505]$.}

\label{Fig2}
\end{center}
\end{figure*}
\\
\begin{figure*}[t]
\centering
  \begin{center}
 \includegraphics[scale=0.5]{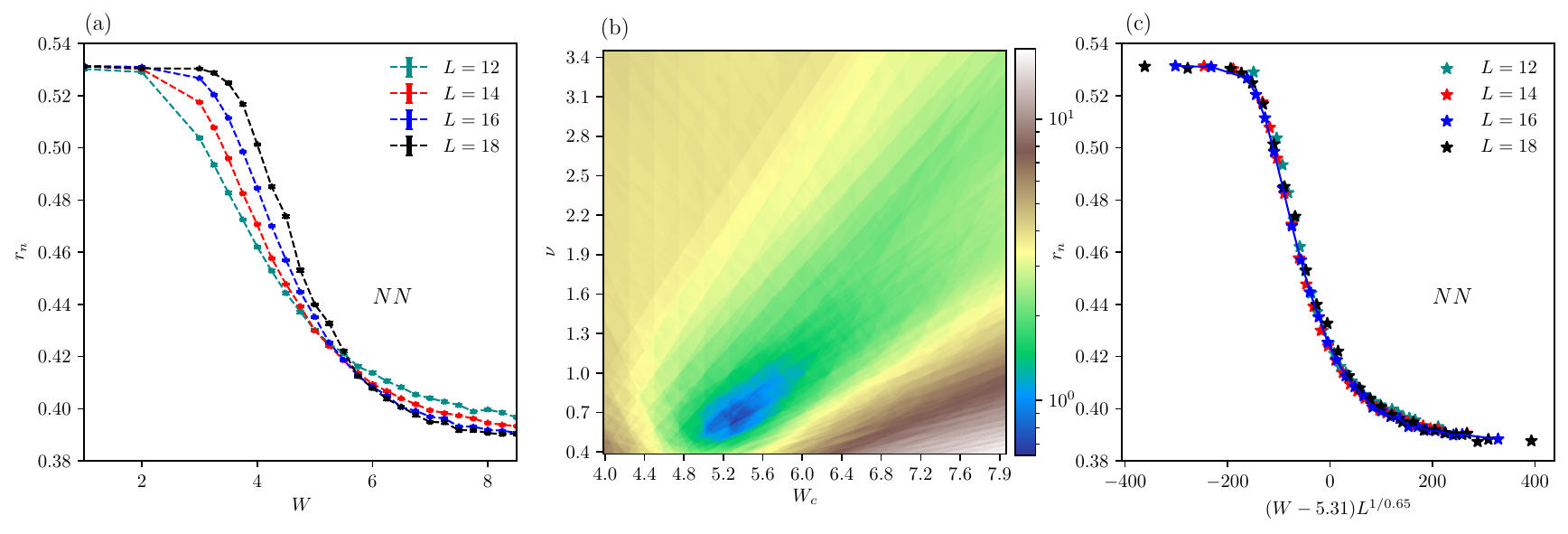}
\caption{Panel(a) shows the level spacing ratio as a function of disorder $W$ for various system sizes for the system with nearest neighbour interactions. Here $r_n$ has been calculated for middle of the many-body eigenspectrum for a rescaled energy bin $\epsilon \in [0.495,0.505]$ . The cost function $C_X$ for the level spacing ratio has been shown in panel (b). The cost function has a minimum for $W_c \sim 5.31$ and $\nu \sim 0.65$. In panel (c) we have shown the scaling collapse using $W_c=5.31t$ and $\nu=0.65$.}
\label{Fig3}
\end{center}
\end{figure*}

\noindent{\bf{Typical LDOS and scattering rates for nearest-neighbour interactions}}:
Further, we analyse the LDOS and scattering rates for the system with nearest neighbour interactions. Fig.~\ref{Fig2} shows the finite size scaling for the ratio of the LDOS and scattering rates.  The cost function for the ratio of the LDOS has a  minima around $W_c \sim 7.96t$ and $\nu \sim 2.76$.
For the ratio of the scattering rates, the cost function has a minima at very close but slightly off values of $W_c \sim 7.81t$ and $\nu \sim 2.32$.
This shows that for all ranges of interactions studied, the LDOS and the scattering rates of the single particle excitations satisfy the CCFS bound. This is because the single-particle excitations are exponentially unlikely to be excited in the MBL phase at large scales though excitations typically propagate up to large length scales in the delocalized phase. This feature of the single particle excitations, and the associated LDOS is basically the property required from a finite volume event in the CCFS argument~\cite{CCFS} to identify the characteristic length $\xi$ and to prove the bound on $\nu$.

Additionally, motivated by renormalization group calculations based on the ``avalanche scenario''~\cite{Vasseur1,Vasseur2,morningstar1}, we also performed the Kosterlitz-Thouless (KT) scaling for all the physical quantities under consideration for various ranges of interactions assuming that the correlation length diverges as $\xi_{KT}=\exp(b/\sqrt{|W-W_c|})$. Based on the calculation of the cost function, we believe that LDOS and scattering rates do not obey the KT scaling as shown in Appendix E. This is indeed expected based on the adiabatic continuity between Anderson insulator and the MBL phase. For Anderson model the typical value of LDOS continuously vanishes at the localization transition point as $(W_c-W)^\beta$ where $\beta$ is proportional to the correlation length critical exponent $\nu$~\cite{Janssen}. One should expect a similar trend of LDOS across the delocalization to MBL transition. Since for the non-interacting Anderson insulator, the correlation length diverges as a power-law with the critical exponent satisfying the CCFS criterion~\cite{Boris,Kramer,Slevin,Biroli,Huse_AM}, one would expect that the same should hold true for the interacting MBL phase. Our numerical analysis of LDOS and scattering rates is consistent with this expectation.
\\
\noindent{\bf{Finite-size scaling of level-spacing ratio}}:
We also analysed the behaviour of the level spacing ratio, which is frequently used to study the MBL transition. The level spacing ratios $r_n$ are defined in the usual way $r_n = \frac{min(\delta_n, \delta_{n+1})}{max(\delta_{n},\delta_{n+1})}, \textrm{~where,~} \delta_n = E_{n+1} - E_n $. Fig.\ref{Fig3} shows the plot of disorder averaged $r_n$ vs disorder $W$ for various system sizes for the system with nearest neighbour interactions. Level spacing ratio obeys Wigner-Dyson statistics for weak disorder and in the very strong disorder limit it obeys the Poissonian statistics. The cost function for level spacing ratio in the $W_c-\nu$ plane has a very different pattern compared to the LDOS and scattering rates studied above. For the level spacing ratio, cost function has a minimum at much smaller value of $W_c^{lsr} \sim 5.31t$ and $\nu \sim 0.65$. With further increase in $W_c$ and $\nu$ the cost function shows a rapid increase. Note that the $W_c$ obtained here is close to the $W_c$ obtained from the KT scaling but the value of the cost-function for the KT scaling is larger indicating better quality of scaling collapse for the power-law diverging correlation length as shown in Appendix E. A similar trend for the cost function of the level spacing ratio is seen for the system with power-law interactions. For all the ranges of interactions studied, we found $\nu <1$. Thus, the critical exponent $\nu$ obtained from the finite-size scaling of the level spacing ratio strongly violates the CCFS criterion, in complete contrast to the LDOS and scattering rates.

Although the ratio of typical to average LDOS and the scattering rate scale with a single parameter such that the critical exponent $\nu \ge 2/d$ with a finite value of the transition point, the level spacing ratio scales with a critical exponent that is much smaller than $2/d$. The most reasonable and physically plausible explanation for this is that different physical quantities approach the thermodynamic limit in different ways. According to our scaling analysis, the transition in level spacing ratio takes place at a disorder value $W_c^{lsr}$ that is smaller than the disorder value at which Green's function quantities undergo transition $W_c$ with $W_C^{lsr} < W_c$. A difference in transition point based on the analysis of different physical quantities has been seen in many earlier works. For example, in some works the transition point from the level spacing ratio was found to be much smaller than that obtained from time evolution of the density imbalance ~\cite{Mirlin_imb} or time evolution of the correlation function and the mean square displacement~\cite{Soumya}. This difference in transition points was explained in terms of the rare region effects which appear in systems with random disorder~\cite{Griffiths}. This is also consistent with recent work~\cite{Huse2022} that proposed various ``landmarks'' between the MBL phase in the thermodynamic limit and the finite-size disordered systems.
However, within our current analysis, one can also not rule out the possibility that the level spacing ratio and single particle excitations may continue to exhibit two distinct transitions even in the thermodynamic limit and if it so the CCFS criterion may not apply at $W_c^{lsr}$ for the transition from ergodic to some intermediate non-ergodic phase because it describes how the correlation length diverges at the transition from the localized to the extended states. But we believe that this is the least plausible scenario.


\section{IV. Conclusions and Discussions}
  The MBL transition involves many higher excited states and entails a transition from the delocalized phase, where eigenstates are extended and obey volume law of entanglement, to the localised side, where eigenstates are localised and obey area law of entanglement. This makes the MBL transition unique and very different from the known transitions in condensed matter systems 
and understanding the nature of the MBL transition is thus central to the problem. 
 We present strong evidence in favour of a continuous delocalization to MBL transition where the correlation length exponent obeys the CCFS criterion.
This is especially significant in light of recent disagreements and controversies regarding the nature of the MBL transition and the stability of the MBL phase. Though for an out-of-equilibrium transition a diverging correlation length may not always be associated with a continuous transition~\cite{Roeck}, the metric we have analysed here, namely, the ratio of typical to average LDOS and scattering rates indeed goes to zero continuously at the MBL transition point and can be used to characterise the delocalization to MBL transition. These results are in striking similarity with the non-interacting Anderson model, where the typical LDOS vanishes at the localization transition point continuously along with the divergent length scale at the transition~\cite{Janssen}.
Our analysis also demonstrates that the MBL phase exists in a system with uniform long-range interactions and nearest neighbour hopping, which is consistent with  existing theoretical~\cite{Burin,Sarang,Gutman,Mirlin,garg_lr,Logan,Bayat} and experimental studies~\cite{zhang_expt,choi}. The MBL transition in systems with uniform long-range interactions is also continuous in nature.

Our findings suggest that there is a strong adiabatic continuity between the interacting MBL phase and the non-interacting Anderson insulator in the strong disorder limit. Regardless of the range of interactions, the strongly disordered interacting MBL phase has single-particle excitations even in highly excited many-body eigenstates not only close to the Fermi energy but even far from it. The concept of adiabatic continuity in disordered interacting systems was first proposed by Anderson~\cite{Fermi_glass} though it was argued  much later that disordered systems with short-range interactions can have localized single-particle excitations~\cite{Basko,Mirlin_0}. Our quantitative findings offer evidence in favour of localised single-particle excitations even in the presence of long-range interacting  MBL phase.

Our numerical analysis demonstrates that the ratio of typical to average LDOS and the scattering rate scale with a single parameter such that the critical exponent $\nu \ge 2/d$ with a finite value of the transition point. In complete contrast to this, conventional diagnostics of the MBL transition, namely, the level spacing ratio scales with a critical exponent that is much smaller than $2/d$. Intriguingly, the exponent obtained from the finite-size scaling of level spacing ratio is quite close to the one obtained from the scaling of the local self energy in the Fock space for the MBL phase~\cite{Logan_scaling}. This may be because the model in Eq.~\ref{Ham} maps onto an effective Anderson model on Fock space and poles of the Fock space propagator are the eigenvalues of the Hamiltonian.  Deep in the localized phase, perturbative corrections to eigenvalues from the hopping terms are directly related to the Feenberg self energy~\cite{Anderson} of the effective Anderson model on Fock space.  Effective Anderson model does not live on a one-dimensional chain but on a complicated Fock graph whose connectivity varies from node to node. For most of the basis states in the middle of the graph, the connectivity scales with physical size of the chain $L$. Thus, it might be possible that the critical exponent of the correlation length in the Fock space obeys a modified generalized CCFS criterion $\nu \ge 4/L$ rather than the standard one, which is written in terms of the physical dimension $d$ of the system. Indeed, the critical exponent obtained from level spacing ratio is also close to the correlation length exponent for the Anderson model on random regular graphs~\cite{Pino,Altshuler}. But this is not the case for the Green's function quantities which involve single-particle excitation energies. This shows that while some physical quantities, like the level spacing ratio, seem to follow the critical exponent of the correlation length in the Fock space, others, like the single particle LDOS studied in this work, and recently explored spatial temperature fluctuations in weakly open MBL systems~\cite{Achim} stick to the system's physical dimension and follow the conventional CCFS bound.

Our work presents a thorough analysis of the universal properties of single-particle excitations across the MBL transition in a class of models with varying range of interaction and provides a clear and strong evidence in favour of a continuous delocalization to MBL transition. Interestingly, both the quantities studied in this work, namely, the single-particle LDOS and scattering rates can be measured in experiments. The search for additional physical quantities that can shed more light on the nature of the MBL transition is unquestionably critical.
\\
\\
\noindent{\bf Acknowledgments}
\\
A. G. would like to thank A. Mirlin and H. R. Krishnamurthy for useful discussions. A.G. acknowledges National Supercomputing Mission (NSM) for providing computing resources of `PARAM Shakti' at IIT Kharagpur, which is implemented by C-DAC and supported by the Ministry of Electronics and Information Technology (MeitY) and Department of Science and Technology (DST), Government of India. V.R.C acknowledges funding from the Department of Atomic Energy, India under the project number 12-R\&D-NIS-5.00-0100.
\\\\

\section{Appendix A: The choice of broadening $\eta$ in the Green's function}

\begin{figure}
      \begin{center}
        \includegraphics[width=3.25in,angle=0]{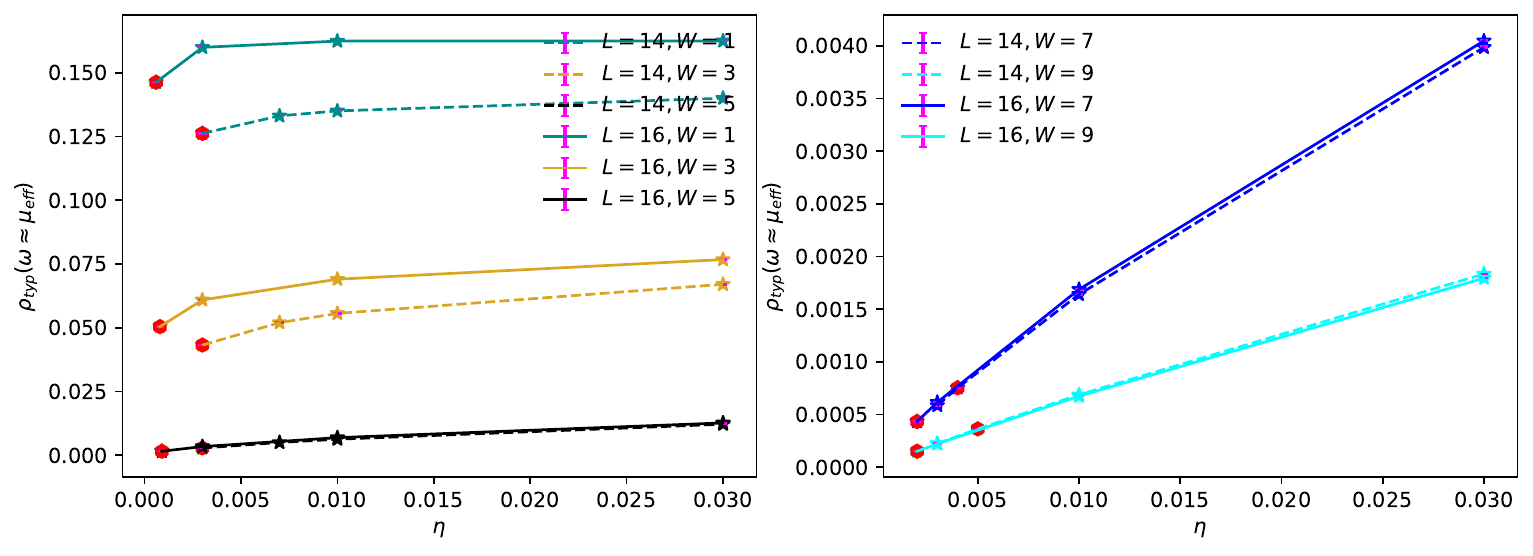}
          \caption{$\rho_{typ}(\omega=\mu_{eff})$ vs $\eta$ for various values of $W$ and $L$. Left panel shows $\rho_{typ}(\omega\sim \mu_{eff})$ vs $\eta$ for weak disorder regime for two system sizes. In the delocalized regime, for any $\eta >0.0075$, $\rho_{typ}(\omega=\mu_{eff})$ is independent of $\eta$. The right panel shows $\rho_{typ}(\omega=\mu_{eff})$ in the localized regime where $\rho_{typ}$ increases monotonically with $\eta$. Thus, $0.0075<\eta<0.035$ provides a legitimate regime of $\eta$ to work with. The data shown is for the case of power-law interactions with $\alpha=1$.}
          \label{rhovseta}
          \end{center}
\end{figure}

  Typical value of LDOS and scattering rate depends upon $\eta$ and how we fix $\eta$ is crucial to the physics of the system. In the case of non-interacting Anderson model, a well defined recipe for appropriate $\eta$ is known as explained in section II. We followed the same method here to fix the broadening for the study of LDOS and scattering rates. We studied $\eta$ dependence of LDOS for a range of disorder values and for various system sizes. Fig.~\ref{rhovseta} shows $\rho_{typ}(\omega=\mu_{eff})$ vs $\eta$ for $L=16$ and $L=14$ and various values of disorder. $\rho_{typ}(\omega=\mu_{eff})$ increases with $\eta$ for strongly disordered phase while in the delocalized phase $\rho_{typ}(\omega=\mu_{eff})$ is almost independent of $\eta$ for $\eta$ between $0.0075$ and $0.03$. Based on this analysis, the infinitesimal $\eta$ is chosen to be $10^{-2}$ in our work for various disorder values and system sizes, where we see $\eta$ independent behaviour of typical DOS in the delocalized phase.

\section{Appendix B: Frequency dependence of LDOS for various ranges of interactions}
\begin{figure*}
\begin{center}
  \includegraphics[scale=0.37]{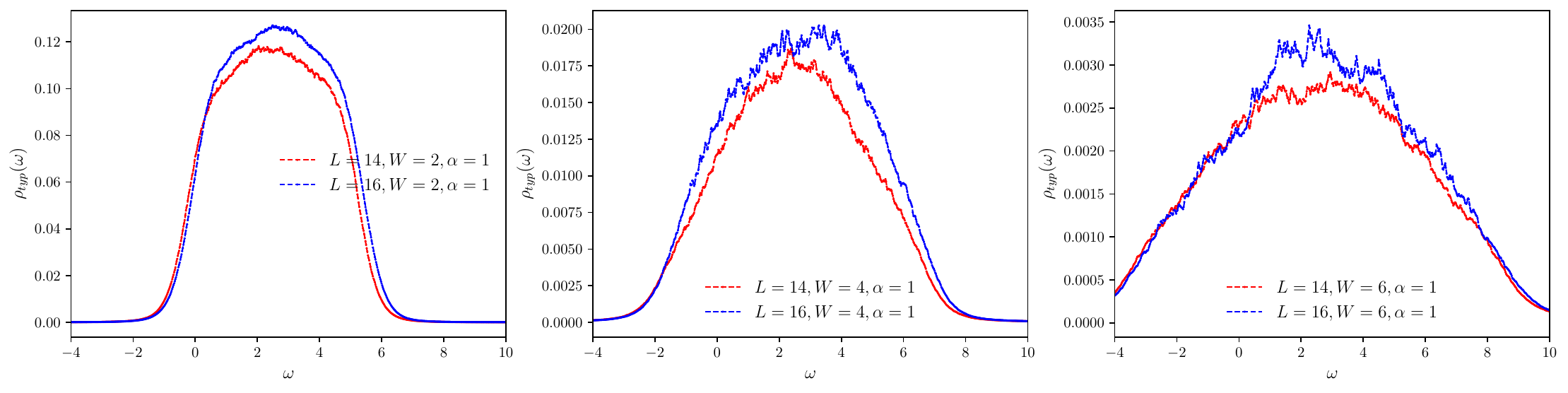}
  \caption{$\rho_{typ}(\omega)$ vs $\omega$ for various values of the disorder strength $W$ for the system with power-law interactions with $\alpha=1$. $\rho_{typ}(\omega)$ is peaked at $\omega\sim \mu_{eff}$ where $\mu_{eff}$ is the chemical potential of the system.}
\label{Fig4_sm} 
\end{center}
\end{figure*}
Fig.~\ref{Fig4_sm} shows the typical value of LDOS $\rho_{typ}(\omega)$ vs $\omega$ for various disorder values and a couple of system sizes. As the
disorder strength increases, $\rho_{typ}(\omega)$ decreases for all the values of $\omega$ and in the MBL phase $\rho_{typ}(\omega)$ becomes vanishingly small for all $\omega$ values. For systems with power-law interactions typical value of local density of states has a peak around $\omega=\mu_{eff}$ where $\mu_{eff}= \sum_{|j-i|=1}^{L/2} \frac{1}{|j-i|^{\alpha}}$ (for $V=1$) is the effective chemical potential of the system under the assumption that the disorder averaged system will respect particle hole symmetry. Thus, for power-law interacting case we have shown scaling of ratio of typical to average value of LDOS and scattering rate for $\omega \sim \mu_{eff}$.

\begin{figure*}
      \begin{center}
        \includegraphics[scale=0.4]{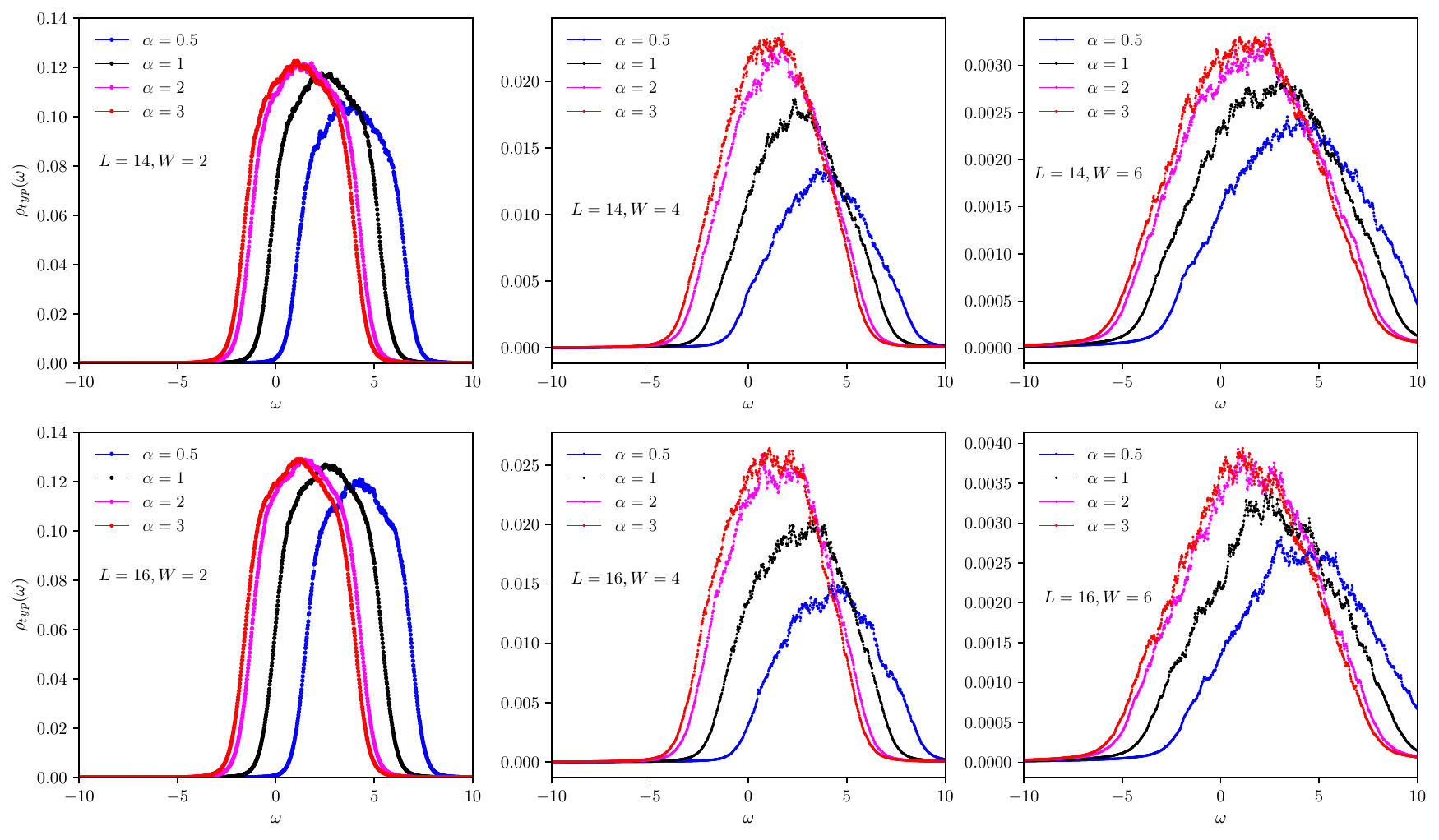}
        \caption{Top panels
          show plots of $\rho_{typ}(\omega)$ vs $\omega$ for various values of disorder,$W$, and for various ranges of interactions for $L=14$. Peak in the $\rho_{typ}(\omega)$ vs $\omega$ curves gets suppressed as the range of interaction increases indicating that $\rho_{typ}(\omega \sim \mu_{eff})$ is smaller for systems with longer range interactions for a fixed disorder strength $W$. Bottom row shows similar data for $L=16$.}
          \label{LR}
          \end{center}
\end{figure*}

In MBL systems, the range of interactions has an intriguing effect on single-particle excitations. Single-particle excitations become more localised for a fixed disorder strength as the range of interactions increases, which lowers the typical values of LDOS. Further, reduced value of typical LDOS implies a weaker scattering among the excitations, longer life-time of excitations and hence a smaller typical value of the scattering rate. Thus, the  effect of disorder is enhanced as the range of interaction increases. This is indeed what is shown in Fig.~\ref{LR} where we have plotted $\rho_{typ}(\omega=\mu_{eff})$  vs $\omega$ for various values of disorder,$W$, and for various ranges of interactions.

\section{Appendix C:  Finite-size scaling of single particle scattering rates}
\begin{figure*}
  \begin{center}
\includegraphics[scale=0.35]{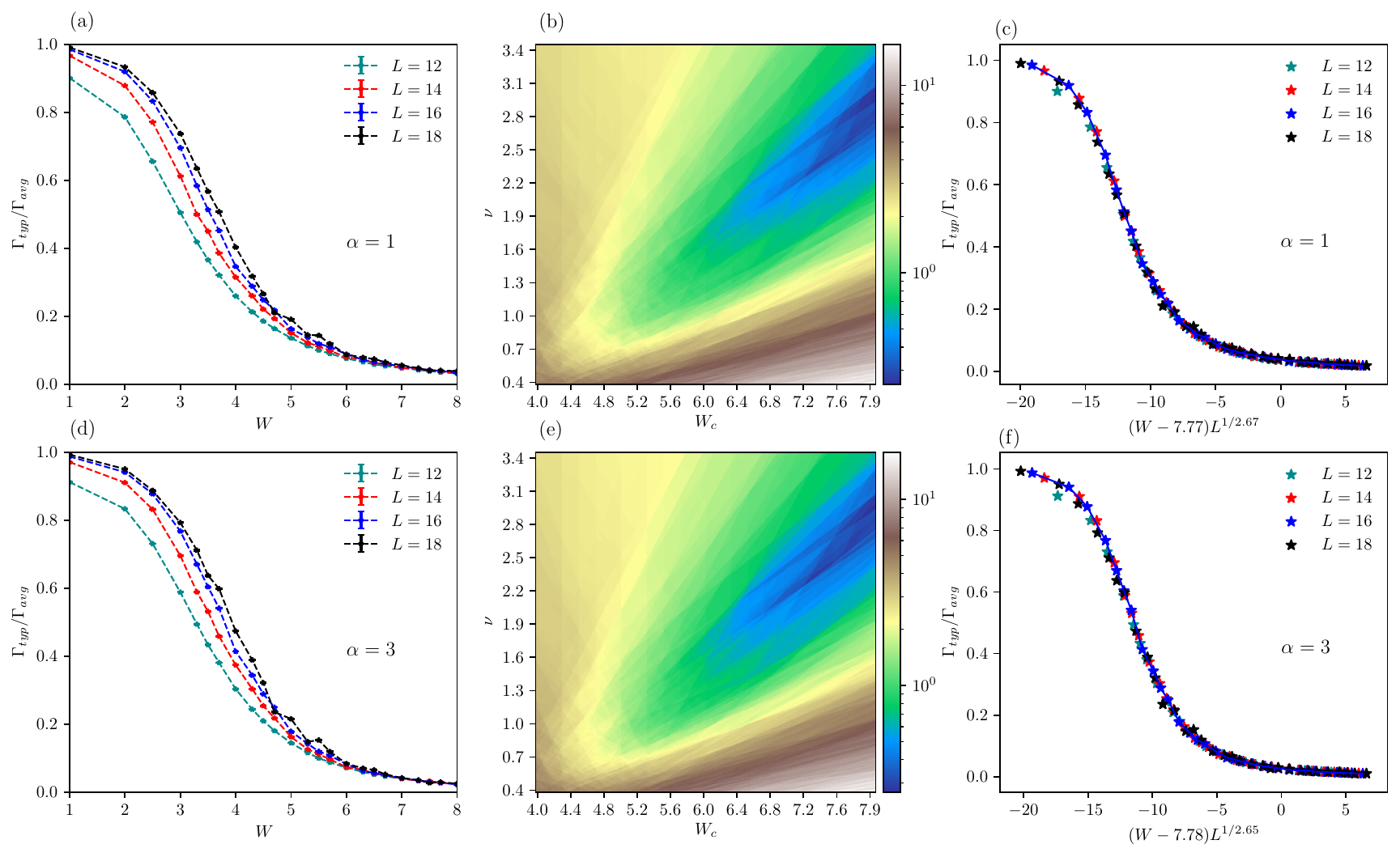}
\caption{{\small{Panel(a) shows the ratio of the typical to average values of the scattering rate $\Gamma_{typ}(\omega)/\Gamma_{avg}(\omega)$ at $\omega=\mu_{eff}$, as a function of disorder $W$ for various system sizes and $\alpha=1$. Panel(b) shows the cost function $C_X$ calculated for $X= \Gamma_{typ}(\omega= \mu_{eff})/\Gamma_{avg}(\omega = \mu_{eff})$ in $W_c-\nu$ plane. The cost function has a minimum  around $W_c \sim 7.77$ and $\nu \sim 2.67$. Panel (c) shows the scaling collapse for $\Gamma_{typ}(\omega=\mu_{eff})/\Gamma_{avg}(\omega = \mu_{eff})$ as a function of the scaled disorder $(W-W_c)L^{1/\nu}$ for $W_c=7.77t$ and $\nu=2.67$. Similar trend of the scattering rates, the corresponding cost function and the scaling collapse is seen for $\alpha=3$ in the bottom row panels. Scattering rate $\Gamma(\omega)$ has been computed for states in the middle of the eigenspectrum for a rescaled energy bin $E \in [0.495,0.505]$. }}}
\label{Fig1_supp}
\end{center}
\end{figure*}

In the main text in Fig. 1 we analyzed the finite-size scaling of single particle LDOS for the system with power-law interactions. We now present details of the finite size scaling for the scattering rates. In \Fig{Fig1_supp}, in the top row we present the data for the system with $\alpha=1$. In the top left panel we show the ratio of typical to average value of the scattering rate $\Gamma_{typ}(\omega)/\Gamma_{avg}(\omega)$ obtained from the middle of the many-body eigenspectrum and $\omega=\mu_{eff}$. In sharp similarity to the LDOS, the ratio of typical to average value of the scattering rate is of order one for weak disorder and becomes vanishingly small and size independent for very large values of disorder. In order to determine the nature of the transition, we did the finite size scaling. As mentioned in the main paper, we calculated the cost function $C_X$ to quantify the finite size scaling collapse. In the top middle panel we show the color plot of the cost function in $W_c-\nu$ plane.  $C_X$ is very large for small values of $W_c$ for any value of $\nu$ considered. For slightly larger values of $W_c$, $C_X$ has a non-monotonic dependence on $\nu$ such that $C_X$ first decrease as $\nu$ increases, attains a minima and then starts increasing again. The best minima obtained in the range of parameters considered, occurs for $W_c \sim 7.77t$ and for $\nu \sim 2.67$. More details about the minimum of the cost function are given in Appendix D. The rightmost panel in the top row shows the scaling collapse as a function of the scaled disorder $(W-W_c)L^{1/\nu}$ with $W_c=7.77t$ and $\nu=2.67$. We would like to emphasize that the ratio of typical to average scattering rate obeys the single parameter scaling and shows a good quality data collapse for the value of the exponent $\nu \ge 2$ which satisfies the CCFS inequality. In the lower row of \Fig{Fig1_supp} we have shown similar plots for $\alpha=3$ which correspond to a shorter range of interactions. As shown here the critical point $W_C$ and the critical exponent $\nu$ are almost independent of the range of interactions. 

\section{Appendix D: Statistical Analysis, error estimation and cost-function minimisation}

In this section, we provide details of standard error on the ratio of typical to average values of the quantities we have studied.
Let $X_i$ with $i=1...N_C$ are the variables for a given set of parameter in the Hamiltonian under study. Here, $i$ corresponds to various sets of $X$ obtained for a large number of many-body eigenstates ($N_E$) lying in the energy bin for which the data has been calculated, the number of lattice sites $L$ and the number of disorder configurations $N_d$ such that $N= N_E \times L \times N_d$. Standard error around the arithmetic mean $X_{avg}$ of $X$ is given by $\Delta X = \frac{\sqrt{ \sum_{i=1}^N (X_i-X_{avg})^2}}{N} = \frac{\sigma}{\sqrt{N}}$. Here $\sigma$ is the standard deviation. Thus, the error-bars around the mean are $X_{avg} \pm \Delta X$.

The geometric standard error around $X_{typ}$ is given by
        \be
        \Delta X_{typ} = \exp\left(\frac{\sqrt{\sum_{i=1}^N (\ln X_i -\ln X_{typ})^2}}{N}\right) 
        \ee
The maximum and minimum values of the typical value $X_{typ}$ are $X_{typ}^{max}=X_{typ} \times \Delta X_{typ}$ to $X_{typ}^{min}=X_{typ}/\Delta X_{typ}$.
Since we are interested in the ratio of typical to average values of the quantity $X_{typ}/X_{avg}$, the range of the ratio is $\left[\frac{X_{typ}^{max}}{X_{avg}-\Delta X}, \frac{X_{typ}^{min}}{X_{avg}+\Delta X}\right]$.The percentage relative error is of order $0.13\%$ for weak disorder and less than $2\%$ for very large disorders for the ratio of typical to average LDOS and scattering rates.
 We further calculated the error bars on the cost function using the error-bars on the ratios of LDOS and scattering rates. Let $X_i$ for $i \in[1,N_{total}]$ are the set of data points used in evaluation of the cost function as in Eq.(2) of the main paper. Statistical error in $C$ is given by  $\Delta C= \frac{\Delta X_1 + 2\sum_{i=2}^{N_{total}-1}\Delta X_{i}+\Delta X_{N_{total}}}{X_{max}-X_{min}} +\frac{(\Delta X_{max}+\Delta X_{min})\sum_{i=1}^{N_{total}}|X_{i+1}-X_{i}|}{(X_{max}-X_{min})^2}+ \Delta C^{(2)}$
  which, upon ignoring the second order term in error, $\Delta C^{(2)}$, gives the relative error in $C$ as $\frac{\Delta C}{C} \sim \frac{\Delta C_{num}}{C_{num}} +\frac{\Delta C_{deno}}{C_{deno}}$ where $C_{num/deno}$ are the numerator or denominator in the Eq.[2] of the main paper.

  So far we have shown density color plots of the cost function,  which is a function of $(W_c,\nu)$. 
  Here we describe how we determine the range of $W_c$ and $\nu$ for which the cost function has a global minimum. We performed a simultaneous minimization of the cost function in $(W_c,\nu)$ plane using a two dimensional sorting algorithm and obtain the transition point and critical exponent corresponding to the minimum $C_{min}$ of the cost-function. Further, we identified the region in $W_c-\nu$ plane corresponding to  minimum of the cost function including statistical error in evaluation of the cost-function $|C_X-C_{min}| \le \Delta C$ where $\Delta C$ is estimate of statistical error in evaluation of the cost-function that we have described above.  We obtained the average value of $W_c$ and $\nu$ as well as the standard error in $W_c$ and $\nu$ over this region. Below we provide details of the average values of transition points and critical exponents along with their standard errors obtained from this procedure for various physical quantities studied.
  
  For the level spacing ratio for the system with nearest neighbour interactions, we found that average value of the transition point $\langle W_c \rangle = 5.31$ and the standard error in the transition point is $\sigma_W = 0.0018$. Similarly, the average value of the critical exponent $\langle \nu \rangle = 0.65$ and the standard error $\sigma_\nu = 0.0013$.  Thus with $95\%$ confidence interval $W_c \in[5.309,5.316]$ and $\nu \in [0.650,0.655]$.
Table~\ref{tab:ldos} presents result of a similar analysis done for the LDOS for various ranges of interactions. 
\begin{table}[htbp] 
  \centering
  \caption{Critical parameters obtained from the finite-size scaling of LDOS $\rho_{typ}/\rho_{avg}$.}
\begin{tabular}
{|c|c|c|c|c|} 
\hline 
Range of interaction & $\langle W_c \rangle $ & $\sigma_W$ & $\langle \nu\rangle$ & $\sigma_\nu$ \\ 
\hline
nearest neighbour & 7.96 & 0.025 & 2.76 & 0.016  \\
\hline
$\alpha=3$ & 7.90 & 0.012 & 2.89 & 0.007 \\
\hline
$\alpha=1$ & 7.67 & 0.016 & 2.62 & 0.012\\
\hline
\end{tabular}

\label{tab:ldos} 
\end{table}
\\
Here $\sigma_W$ is the standard error in estimation of $W_c$ and $\sigma_\nu$ is the standard error in $\nu$. Thus, with $95\%$ confidence interval $\nu \in [2.729,2.793]$ for nearest neighbour case and also for power-law interactions $\nu >2$ with a higher than $99\%$ confidence interval. For the ratio of typical to average value of scattering rates, analysis of the cost-function gives following values for $W_c$ and $\nu$:
\begin{table}[htbp] 
\centering 
\caption{Critical parameters obtained from the finite-size scaling of $\Gamma_{typ}/\Gamma_{avg}$.} 
\begin{tabular}
{|c|c|c|c|c|} 
\hline 
Range of interaction & $\langle W_c \rangle $ & $\sigma_W$ & $\langle \nu\rangle$ & $\sigma_\nu$ \\ 
\hline
nearest neighbour & 7.80 & 0.021 & 2.32 & 0.012 \\
\hline
$\alpha=3$ & 7.78 & 0.022 & 2.65 & 0.015 \\
\hline
$\alpha=1$ & 7.77 & 0.042 & 2.67 & 0.010\\
\hline
\end{tabular} 
\label{tab:gamma} 
\end{table}
\\
Finite-size scaling of the scattering rate ratios also shows that the critical exponent $\nu >2$ with higher than $99\%$ confidence interval for all the ranges of interactions studied. 
\\\\
 \begin{figure*}[t]
\begin{center}
\includegraphics[width=0.7\textwidth]{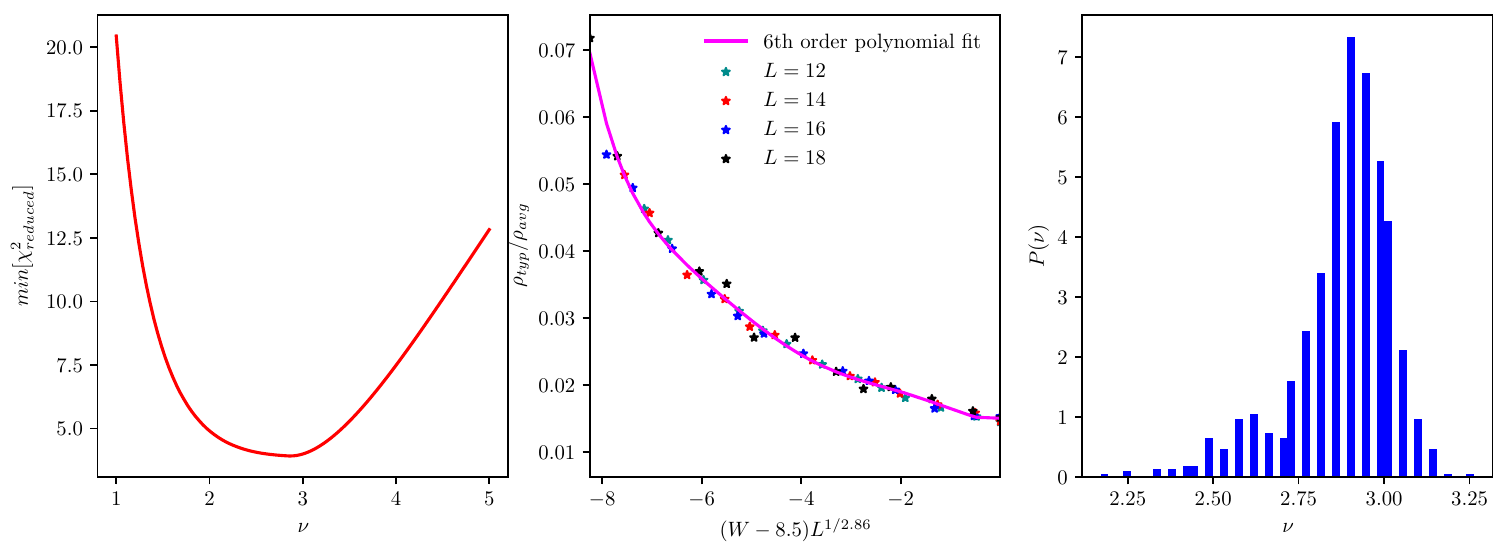}
\caption{The left panel shows the behaviour of the minimum of $\chi^2_{reduced}$ of the polynomial
fit of the scaling function (we use a sixth order polynomial) as a function of the correlation length exponent for a typical
bootstrap sample. The middle panel shows the data collapse using $W_c, \nu$ at the minimum of $\chi^2_{reduced}$ in the left panel
which are the critical parameters for this sample.
The right panel
depicts the probability density of $\nu$ values produced by using $1000$ bootstrap samples. We use $\alpha=1$ and
$W \in [5.5, 8.5]$ for this figure.
}
\label{cnf_int}
\end{center}
\end{figure*}
{\bf{Confidence intervals based on resampling: }} Till now in this Appendix we have provided error estimates of the critical parameters using error propagation analysis of the cost function. We now provide an alternative evaluation of confidence intervals for the same parameters using a $\chi^2$ analysis of our scaling collapse and resampling of our data in the spirit of a bootstrap analysis. \\

We use the cost function analysis we have done till now to first identify the likely location of a critical region. For each lattice size and
the relevant disorder strength for that region, we evaluate statistically independent values of the order parameter by using a number of subsets
of the available disorder realisations. This provides us with a value and error estimate of the order parameter. We then fit the scaling collapse
plot for this data to a polynomial function and identify the best collapse by minimizing the reduced $\chi^2$ for the fit,  
$\chi^2_{reduced} \equiv \frac{1}{N_{deg}}\sum_i \frac{(Y_i - y_i)^2}{{\delta y_i}^2} $ as a function of $W_c, \nu$. Here $y_i$ and $\delta y_i$ 
are the original data and the corresponding errors and $Y_i$ is the value obtained from the fit and $N_{deg}$ the number of degrees of freedom for the fit. This minimisation gives us the values of $(W_c, \nu)$ for this sample. 
We repeat this procedure multiple times (i.e. generate many bootstrap samples) to get independent evaluations of the critical parameters and 
thence their errors and confidence intervals. Fig. \ref{cnf_int} depicts typical results of such and analysis for $\alpha=1.0$.

For the depicted case we used the data for all the lengths with disorder strengths in the range of
$W \in [5.5, 8.5]$ for $\alpha=1$. For a given bootstrap sample the available disorder realisations are divided
into $10$ blocks for $L=12, 14,16$ and into $5$ blocks for $L=18$. We use a $1000$ bootstrap samples for evaluation
of the critical parameters and fit the scaling function for each sample to a polynomial of order $6$.
The mean and standard deviation of the evaluated $\nu$ values are $[2.88, 0.15]$ and
the same for the $W_c$ values are $[8.45, 0.1]$. As is clear from the figure and these numbers,
the value of $\nu$ is greater than $2$ with a higher than $99\%$ confidence interval. 

As of now we cannot rule out the possibility that the critical point might be further away from the
largest disorder strength that we have studied. In line with this, we note that these values of critical
parameters do change somewhat if we choose a different critical region, a different order for the polynomial fit
etc. However, in all such cases we have analysed we always get $\nu$ to be greater than $2$ (with
similar accuracy) as we found above. So, while a very accurate determination of the critical disorder
strength or a very precise value of the correlation exponent will certainly need a much more extensive analysis
with more lattice sizes and disorder strengths, we expect the $\nu>2$ result to be robust and to hold quite generally.

  \section{Appendix E: Kosterlitz-Thouless Scaling}
  \begin{figure*}
  \begin{center}
     \includegraphics[width=4.0in,angle=0]{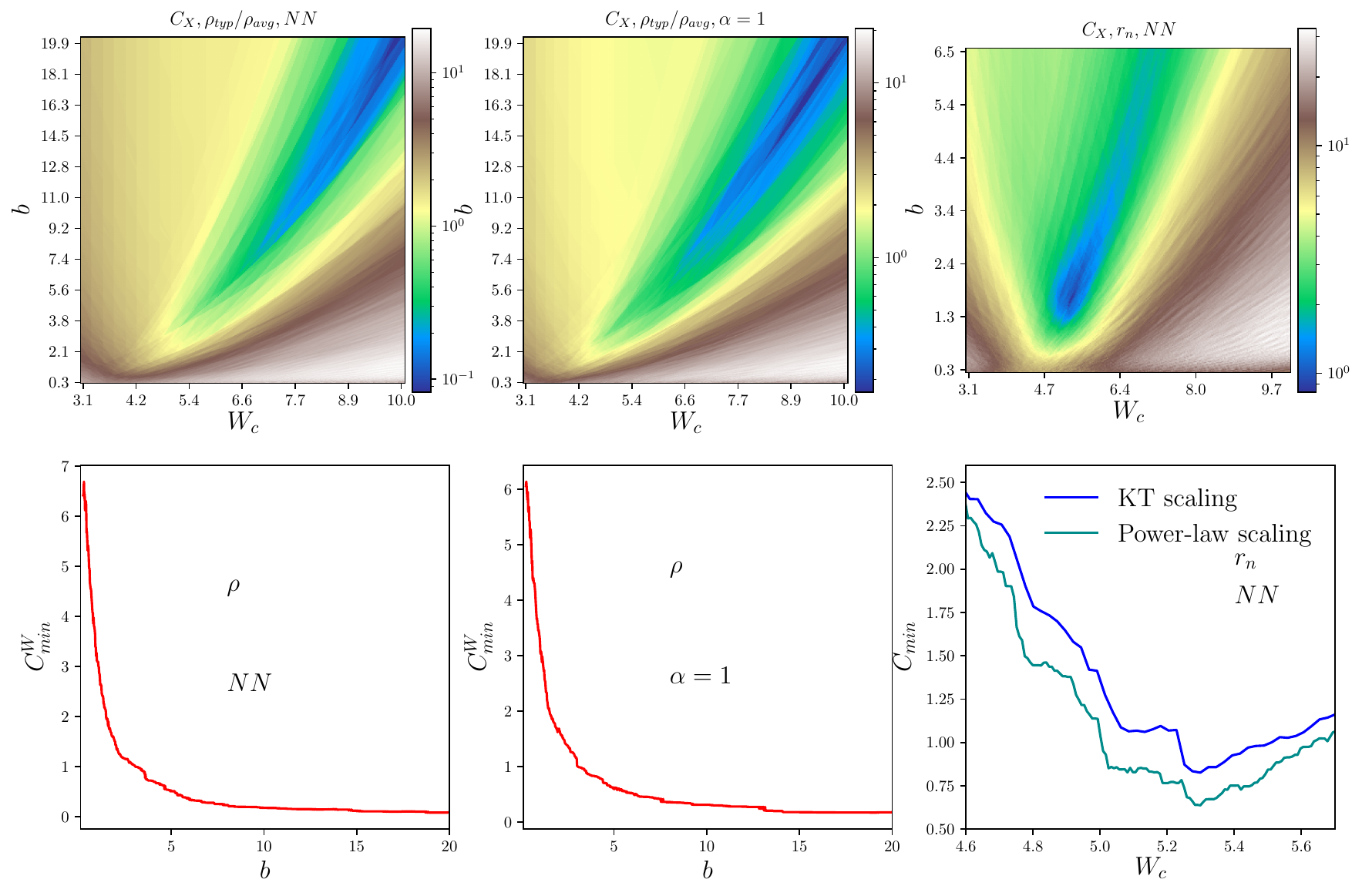}
        \caption{Top row shows the cost function color-plots in the $(b-W_c)$ plane. The  first and the second plot in the top row are the cost functions for the  KT scaling of the ratio of typical to average density of states for systems with nearest neighbour interactions and power-law interactions with $\alpha=1$ respectively. The corresponding cost-function minimum $C^W_{min}$ vs $b$ is shown in the bottom layer. There is no finite value of $b$ of order unity for which the cost function attains a minima, rather $C^W_{min}$ keeps decreasing slowly as $b$ is increased all the way upto very large values like $b=20$. The third panel shows the cost function for the level spacing ratio for the system with nearest neighbour interactions, for which the transition point $W_c\sim5.3t$ is very close to the one obtained from power-law divergence of the level spacing ratio but the minimum of the cost function, which is at $b=1.6$, is more than that for the power-law diverging correlation length as shown in the 3rd bottom panel.} 
          \label{KT}
          \end{center}
\end{figure*}
  In this section we present results for the Kosterlitz-Thouless (KT) scaling of various physical quantities that we have explored in this work. We assume that the correlation length diverges as
    \be
    \xi_{KT} \sim exp(\frac{b}{\sqrt{|W-W_c|}})
    \ee
    We arranged the values for ratio of observable $X_i$ according to increasing values of $s \frac{L}{\xi_{KT}}$ where $s=sgn(W-W_c)$. The arranged values of $X_i$ are used to evaluate the cost function defined in Eq.(2) of the manuscript. Fig.~\ref{KT} shows the color-plots of the cost function for the ratio of typical to average value of LDOS in the $(b,W_c)$ plane. Ideally parameter $b$ should be of order unity but for no finite small value of the parameter $b$, we could find a minima of the cost function for any range of interaction as shown in the Fig.~\ref{KT}. The cost function keeps decreasing slowly as $b$ increases all the way up to $b=20$ or so. We believe that this is an indication of the fact that KT scaling does not work properly for the quantities like LDOS and scattering rates. This analysis shows that though for level spacing ratio both the KT scaling and power-law diverging correlation length ansatz seems to provide a reasonably good scaling, albeit the minimum of the cost function being higher for the KT scaling; quantities describing single-particle excitations in the systems do not obey the KT scaling ansatz further confirming our claim that single-particle excitations contain signature of a continuous delocalization to MBL transition which satisfies the CCFS criterion.
For level spacing ratio, we do find a transition point $W_c \sim  5.3$ from the KT scaling which is close to the one obtained from the finite-size scaling assuming a power-law diverging correlation length with the cost function minimum occurring at $b\sim 1.6$. But the minimum value of the cost function is higher than that for the power-law diverging correlation length indicating that the power-law divergence provides a better scaling collapse for the level spacing ratio as shown in 3rd panel of bottom row of Fig~\ref{KT}.

\begin{thebibliography}{99}
\bibitem{footnote}
   This is a much stronger criterion than the original one established by Harris~\cite{Harris} according to which if $\nu \ge 2/d$ for a clean system undergoing a continuous transition then disorder is irrelevant in the renormalisation group sense.
\end{thebibliography}
\end{document}